\documentclass[12pt,usenatbib,iop]{emulateapj}
\newcommand       \erg		{\,{\rm erg }}

\newcommand       \sinvert		{\,{\rm s$^{-1}$ }}

\newcommand       \HII          {\,H\,{\footnotesize II} }
\newcommand      \calU       {\mathcal{U}}
\newcommand      \rch         {\,\tilde{r}_{\rm ch}} 
   
\newcommand      \OIII        {\,O\,{\footnotesize III}\,}
\newcommand      \NII        {\,N\,{\footnotesize II}\,}

\newcommand     \stromgren  {{Str\"{o}mgren} }

\begin{document}

\slugcomment{Accepted for publication in ApJ, March 12, 2013}

\title{Line Emission from Radiation-Pressurized HII Regions\\I: Internal Structure and Line Ratios} 
\author{Sherry C. C. Yeh\altaffilmark{1}, Silvia Verdolini\altaffilmark{2}, Mark R. Krumholz\altaffilmark{3}, Christopher D. Matzner\altaffilmark{1}, Alexander G. G. M. Tielens\altaffilmark{2}}

\email{yeh@astro.utoronto.ca}

\altaffiltext{1}{Department of Astronomy \& Astrophysics, University of Toronto, 50 St. George St., Toronto, ON M5S 3H4, Canada}

\altaffiltext{2}{Leiden Observatory, University of Leiden, P. O. Box 9513, 2300 RA Leiden, Netherlands}

\altaffiltext{3}{Department of Astronomy and Astrophysics, University of California, Santa Cruz, CA 95064}

\begin{abstract}
The emission line ratios [\OIII]$\,{\lambda 5007}$/H$\beta$ and [\NII]$\,{\lambda 6584}$/H$\alpha$ have been adopted as an empirical way to distinguish between the fundamentally different mechanisms of ionization in emission-line galaxies. However, detailed interpretation of these diagnostics requires calculations of the internal structure of the emitting \HII regions, and these calculations depend on the assumptions one makes about the relative importance of radiation pressure and stellar winds. In this paper we construct a grid of quasi-static \HII region models to explore how choices about these parameters alter \HII regions' emission line ratios. We find that, when radiation pressure is included in our models, \HII regions reach a saturation point beyond which further increases in the luminosity of the driving stars does not produce any further increase in effective ionization parameter, and thus does not yield any further alteration in an \HII region's line ratio. We also show that, if stellar winds are assumed to be strong, the maximum possible ionization parameter is quite low. As a result of this effect, it is inconsistent to simultaneously assume that \HII regions are wind-blown bubbles and that they have high ionization parameters; some popular \HII region models suffer from this inconsistency. Our work in this paper provides a foundation for a companion paper in which we embed the model grids we compute here within a population synthesis code that enables us to compute the integrated line emission from galactic populations of \HII regions.
\end{abstract}
 

\section{Introduction}

The line ratios [\OIII]$\,{\lambda 5007}$/H$\beta$ and [\NII]$\,{\lambda 6584}$/H$\alpha$, first proposed for use in galaxy classification by \citet{bpt81} (hereafter BPT), are commonly used to diagnose the origins of emission lines from galaxies, and in particular to discriminate between galaxies whose emission is powered by star formation-driven \HII regions and from those powered by active galactic nuclei (AGNs). These emission line pairs are particularly useful because (1) they are bright and thus relatively easy to measure, (2) blending between the lines  can be corrected with reasonable accuracy, so long as the spectra are taken with sufficient resolution,  
and (3) the wavelengths in each line pair are quite similar, so the line ratio is relatively insensitive to dust-reddening \citep{veilleux87}. 

The power of these line ratios as diagnostics comes from their sensitivity to the spectral shape of the radiation field driving the ionization, which can be understood from a simple physical picture. To first order, the intensities of the H$\alpha$ and H$\beta$ lines simply measure the total photoionization rate, and thus normalize out the ionizing luminosity. On the other hand, the [\OIII] and [\NII] intensities are sensitive not only to the total ionizing luminosity, but also to the shape of the ionizing spectrum and to the ionization parameter $\calU$, which measures the ratio of photons to baryons in the ionized gas. When the ionizing flux arises from hot stars, the ionizing spectrum is dominated by low-energy photons that have short mean-free paths through neutral gas. Thus the \HII region consists of a fully ionized zone with a sharp boundary. Within this region, as $\calU$ increases, more of the ionized gas volume becomes filled with high ionization-potential species such as O$^{++}$, and less with low-ionization potential species such as N$^+$. As a result, \HII regions ionized by hot stars tend to fall along a sequence that runs from high [\OIII]/H$\beta$, low [\NII]/H$\alpha$ to low [\OIII]/H$\beta$, high [\NII]/H$\alpha$. On the other hand, if the ionizing spectrum follows a power-law, as expected for AGN, then a significant amount of the ionization is produced by X-ray photons capable of ionizing higher ionization potential species like O$^{++}$. Moreover, these photons have large mean-free paths, giving rise to a large zone of partial ionization rather than a smaller region of full ionization as in the stellar case. In this configuration, [\OIII]/H$\beta$ and [\NII]/H$\alpha$ both increase with $\calU$, and either one or the other tends to be larger than in the stellar case, leading to a sequence that runs from intermediate to high [\OIII]/H$\beta$ and [\NII]/H$\alpha$ and is well-separated from the locus occupied by \HII regions dominated by stellar sources.

This simple picture is roughly consistent with local observations:
star-forming galaxies in the Sloan Digital Sky Survey (SDSS; median redshift $z=0.1$) obey a tight correlation between the [\OIII]/H$\beta$ and [\NII]/H$\alpha$ ratios in the BPT diagram  \citep{brinchmann04a, tremonti04a}.  However, higher redshift star-forming galaxies are offset from this sequence to higher 
values of 
[\OIII]/H$\beta$, 
without joining the 
locus of points occupied by AGN in the SDSS sample \citep{shapley05, erb06, erb10, brinchmann08, liu08}.  
 As our work is motivated by a desire to better understand the physical information encoded in the BPT diagram, we pause to consider how this shift in the BPT locus might arise.  

The difference in line luminosity ratios could be intrinsic to the galaxies' \HII regions.  As we have said, the ionization parameter is a major controlling factor which positions regions {\em along} the star-forming locus.  Line emission from these regions will be also affected by the metallicity and dust content of the interstellar gas; by the density of that gas (through the critical densities of the lines); by the ionizing spectra of the stars (which reflect stellar masses, metallicities and rotation rates).    In addition, stellar winds can alter the boundary conditions for ionized zones, a point discussed by \citeauthor{yeh12} (\citeyear{yeh12}, hereafter YM12) and to which we return below. 

Alternatively, the shift in the BPT diagram could arise from outside the \HII regions if their light is mixed with line emission from shocks or an unresolved AGN (e.g., \citealt{liu08}).    Indeed, \citet{Wright10} use integral field spectroscopy to demonstrate that a weak AGN is responsible for the shift in a single galaxy at $z=1.6$, and \citet{Trump11} stack HST grism data of many galaxies to show that this phenomenon is reasonably common.   Taken together, these studies raise the possibility that \HII regions at $z\approx2$ lie along the same BPT locus as those nearby, and the shift is an optical illusion caused by active nuclei.   

However, the distribution of high-redshift galaxies in the BPT diagram is {\em also} shifted in the direction of high $\calU$.    Because radiation pressure rises, relative to gas pressure, in proportion to $\calU$, this implies that the radiation force typically is more important in high-redshift galaxies.  
This radiation-force-dominated condition is also more prevalent among starburst galaxies in the local Universe, as 
YM12 argue on the basis of mid-infrared line emission. 
This possibility has also received significant support from recent resolved observations of \HII regions,
which provide 
direct evidence that radiation pressure is significant for the most luminous examples
( \citealt{lopez11a}, but also see \citealt{pellegrini11a} and \citealt{silich13}).

The detailed role of radiation pressure in altering the line ratios of starlight-ionized \HII regions has received relatively little attention, although the phenomenon has been explored in the context of AGN narrow-line regions \citep{Binette97,Dopita02}. Early models ignored radiation pressure entirely \citep[e.g.][hereafter D00]{dopita00}.  Although more recent models include radiation pressure \citep[e.g.][]{dopita05, dopita06, dopita06b, groves08a, levesque10}, it is   
either explicitly or implicitly assumed that the geometries and internal structures of \HII regions are dominated by stellar wind bubbles rather than radiation pressure. 
As we discuss below, the assumption that stellar wind pressure exceeds radiation pressure is often physically inconsistent
with the range of ionization parameters being probed.  
Moreover, resolved observations of the brightest nearby \HII regions indicate
the hot gas produced by shocked stellar winds for the most part does not remain confined within \HII regions, and instead leaks out into the low-density ISM \citep{townsley03a, harperclark09}. As a result, the pressure of shocked stellar wind gas is often smaller rather than larger than radiation pressure (\citealt{lopez11a}; YM12)\footnote{
Note that \citet{pellegrini11a} assume a smaller filling factor for the X-ray emitting gas, and so assign it a much higher pressure than \citet{lopez11a}. 
 For the same luminosity, small, higher-pressure bubbles have a greater dynamical effect on their immediate surroundings. 
 But these bubbles are less important for the entire region than large, lower-pressure ones.  
 This is a consequence of the virial theorem, which ties dynamics to the net energy budget.
}.

In this paper we explore how radiation pressure influences the line emission of \HII regions. To do so, we compute a sequence of hydrostatic \HII region models under a variety of physical assumptions about the relative importance of radiation pressure and stellar wind pressure (\S \ref{S:static_models}), and we explore how varying the physical assumptions alters the loci occupied by the model \HII regions in the BPT diagram (\S \ref{S:model_results}). We then compare our models to those published by other authors (\S \ref{sec:model_comparison}) and draw conclusions (\S \ref{S:BPT_discussion}). In a companion paper (Verdolini et al.~2012, hereafter Paper II), we use the grid of \HII region models presented in this paper to construct a population synthesis model capable of predicting the line ratios of star-forming galaxies containing many different \HII regions.  We use these models to compare to observations of star-forming galaxies. Although this study cannot replace a full investigation of the factors affecting the \HII region locus within the BPT diagram, it is the first to explore the roles of radiation and wind pressure. 

\section{Photoionization Models}\label{S:static_models}

\subsection{Input Parameters and Calculations}

\begin{deluxetable*}{lccc}
\tabletypesize{\scriptsize}
\tablewidth{0pt}
\tablecaption{Comparison of Model Parameters}
\tablehead{
\colhead{~} &
\colhead{\citealt{dopita00} Model} &
\colhead{Our Model} &
\colhead{\citealt{levesque10} Model}
}
\startdata
Ionizing Spectra & Starburst99 & Starburst99 & Starburst99\\ 
Initial Mass Function & Salpeter, M$_{\rm up} =$\,120 M$_{\sun}$ & Default Starburst99 IMF\tablenotemark{a}, M$_{\rm up} =$\,120 M$_{\sun}$ & Salpeter, M$_{\rm up} =$\,100 M$_{\sun}$\\
Stellar Tracks & Geneva standard & Geneva standard & Geneva high mass-loss \\
Stellar Atmospheres & Lejeune-Schmutz & Lejeune-Schmutz & Pauldrach/Hillier \\ 
Photoionization Code & MAPPINGS III & Cloudy v08.00 & MAPPINGS III\\
Radiation Pressure & Not included & Optional & Included\\
Geometry & Plane-Parallel & Spherical & Plane-Parallel\\
H~\textsc{ii} Region State & Uniform gas pressure & Hydrostatic & Isobaric\tablenotemark{b}\\
Metallicity & Solar & Solar & Solar \\
\cutinhead{Gas Phase Abundances (logarithmic, relative to H)}
H &	0  & 0 & 0 \\
He &	-1.01& -1.01 & -1.01 \\
C &	-3.74 & -3.74  & -3.70 \\
N &	-4.17 & -4.17 & -4.22 \\
O &	-3.29 & -3.29 & -3.29 \\
Ne &  -3.91& -3.91 & -3.91 \\
Mg &  -5.12 &- 5.12 & -5.12 \\
Si & 	-5.45 & -5.45 & -5.45 \\
S &	-4.79 & -4.79 & -5.19 \\
Ar &	-5.44 & -5.44 & -5.44 \\
Ca &	 -8.16 & -8.16 & -8.16 \\
Fe &	-6.33 & -6.33 & -6.33
\enddata
\tablenotetext{a}{IMF exponents: 1.3 and 2.3 at mass boundaries 0.1, 0.5, and 120M$_\sun$.}
\tablenotetext{b}{{L10} included radiation pressure in their isobaric calculations, thus the hydrostatic condition reduces to a state of uniform total pressure in plane symmetry (YM12). } 
\label{tab:model_parameters}
\end{deluxetable*}

To study the influence of radiation pressure on \HII regions, we construct a grid of static, single \HII regions, with a wide range of sizes, ionizing luminosities, and wind strengths. Our procedure is as follows. We first use the stellar population synthesis code Starburst99 \citep{sb99} to generate spectra from coeval star clusters. We assume that all of the clusters are massive enough to fully sample the stellar initial mass function, which we take to have exponents -1.3 and -2.3 between stellar mass boundaries 0.1, 0.5, and 120 M$_{\sun}$. We employed the Geneva standard evolutionary tracks \citep{char96,schaerer93b,schaerer93a,
schaller92} with solar metallicity, and Lejeune-Schmutz stellar atmospheres \citep{lejeune97,lejeune98,schmutz98}, which incorporate plane-parallel atmospheres and stars with strong winds. We record the Starburst99 output spectra for cluster ages of 0 to 11 Myr at 0.5 Myr intervals.

We then use the photoionization code Cloudy 
08.00, 
last described by \citet{cloudy98}, to compute the structure of static, spherical \HII\ regions driven by point sources whose spectra are taken from the Starburst99 calculations. In addition to the spectrum of the driving source, Cloudy requires a number of other input parameters. The first of these is the total luminosity of the ionizing source, for which we run a series of models with  $L=10^{33} - 10^{46}$ \erg \sinvert in 1 dex steps. The second is the number density of hydrogen nuclei at the innermost zone of the \HII region, which we set to values from $n_{\rm H, in} = 10^{-1} - 10^5$ in steps of 1 dex.  The third is the distance of the innermost zone from the point source, which we vary from $R_{\rm in} = 10^{-2} \rch - 10^2 \rch$ in steps of 0.2 dex. Here the characteristic radius 
\begin{equation}
\label{eq:rch}
\rch = \frac{\alpha L^2}{12\pi (2.2 k_B T c)^2 S}
\end{equation}
is the radius of a uniform-density, dust-free Str\"omgren sphere for which the gas pressure is equal to the unattenuated radiation pressure at its edge (\citealt{krumholz09d}; YM12); in this equation, $\alpha$ is the recombination rate coefficient, $T$ is the gas temperature, and $L$ and $S$ are the bolometric luminosity and the output of ionizing photons per unit time, respectively, from the point source. In all the calculations presented here we adopt the same fiducial parameters as YM12: $T = 8000$ K, $\alpha = 3.0 \times 10^{-13}$ cm$^3$ s$^{-1}$. Finally, we adopt Cloudy's default ISM dust grain abundance and size distributions at solar metallicity, but in order to allow more meaningful comparison between our results and those of D00, we adjust the
 gas-phase
element abundances in our calculation to match theirs. 
These choices mean that the dust discriminant parameter \citep{Draine11} takes the same value, $\gamma=7.4$, as in YM12: so, dust opacity is significant within radiation pressure-dominated ionized zones. 
We summarize all the parameters we use in our calculations in Table~\ref{tab:model_parameters}. The Table also describes the parameter choices used in D00 and \citet[hereafter L10]{levesque10}.

For each set of input parameters, we use Cloudy to calculate the structure of the resulting \HII region, halting at the point where the gas temperature drops to 100 K in order to ensure that the ionization front is 
fully enclosed. 
We perform each calculation twice, once with radiation pressure turned off, and once with radiation pressure turned on and allowed to exceed gas pressure (in contrast to Cloudy's default setting, which does not allow radiation pressure to be greater than gas pressure.)

\subsection{Model Outputs and Physical Parameters}
\label{SS:model_outputs}

The output of our calculations is two four-dimensional grids of models defined by the parameters $(t, n_{\rm H,in}, L, R_{\rm in})$, where $t$ is the age of the stellar population used to generate the ionizing spectrum. One grid contains models with radiation pressure turned on, which we refer to as the RP models, and the other contains models with radiation pressure disabled, which we refer to as gas pressure, or GP, models. For each run in both model grids, we compute several optical emission line luminosities integrated over the ionized gas, including the lines used to construct the BPT diagram: H$\alpha$, H$\beta$, [\OIII]$\,{\lambda 5007}$, and [\NII]$\,{\lambda 6584}$.

In order to understand the physical meaning of the results, it is helpful to characterize each model by two dimensionless numbers that can be computed from the 
Cloudy
output.
Following YM12, we define the radiation pressure parameter
\begin{equation}\label{eq:Psi}
\Psi \equiv {R_{\rm IF}\over \rch},
\end{equation} 
where $R_{\rm IF}$ is the radius of the ionization front (IF). A value of $\Psi < 1$ indicates that the entire IF falls within $\rch$, and thus that radiation pressure is more important than gas pressure in determining its structure. Again following YM12, we define a separate stellar wind parameter
\begin{equation}
\Omega \equiv \frac{P_{\rm in}V_{\rm in}}{P_{\rm IF}V_{\rm IF} - P_{\rm in}V_{\rm in}},
\end{equation}
where $P_{\rm IF}$ and $P_{\rm in}$ are the gas pressures at the edge of the ionization front and the innermost zone, respectively, and $V_{\rm IF}=(4/3)\pi R_{\rm IF}^3$ and $V_{\rm in}=(4/3)\pi R_{\rm in}^3$ are the volumes contained within the IF and the inner edge of the \HII region, respectively. The inner edge of the photoionized region is the outer edge of the bubble of hot gas inflated by the stars' winds. Thus $\Omega$ reflects the contribution of a pressurized wind bubble to the total energy budget of the \HII region. In a region strongly pressurized by stellar winds $\Omega \gg 1$, while in a region with negligible wind pressure $\Omega \ll 1$.

\begin{figure}
\epsscale{1}
\plotone{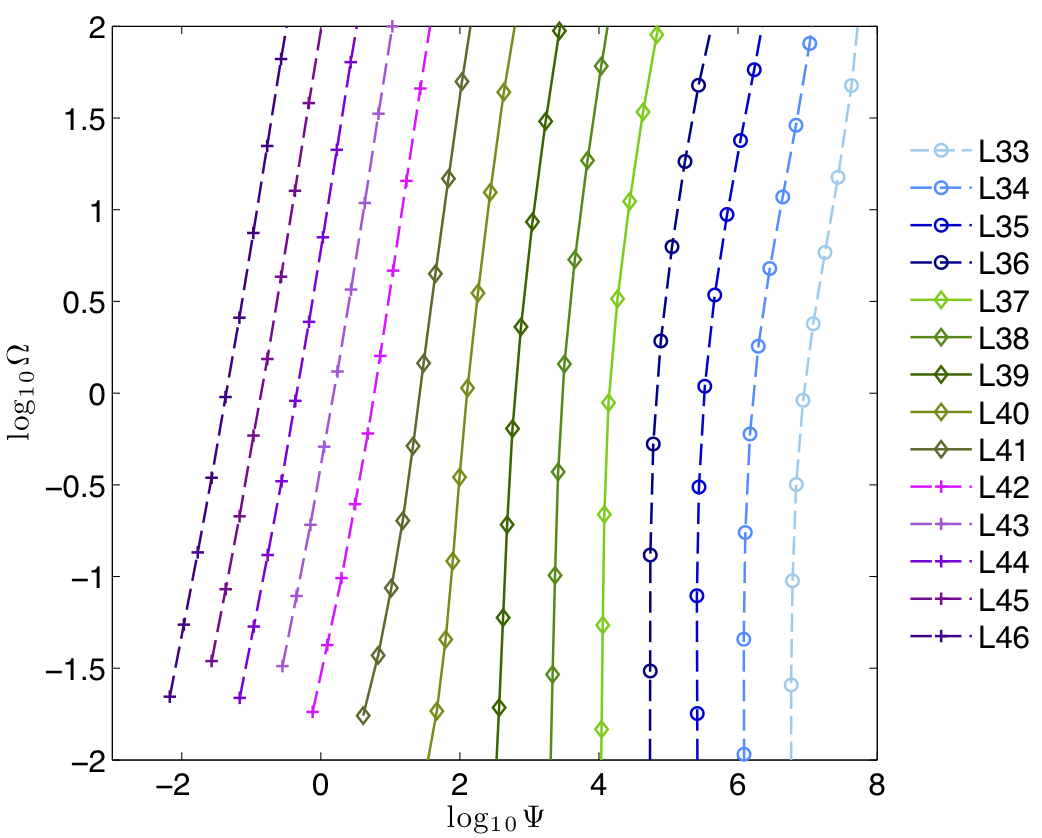}
\caption{
\label{fig:param_example}
An example of how our models fill the parameter space of $\Psi$ and $\Omega$. In the Figure, each plot symbol shows the values of $\Psi$ and $\Omega$ computed for a particular calculation in our model grid at $t=0$, $n_{\rm H,in} = 10$ cm$^{-3}$, with radiation pressure on. Colors indicate lines of constant $L$, running from $10^{33} - 10^{46}$ erg s$^{-1}$ as indicated in the legend. The sequence of points along a given model corresponds to varying $R_{\rm in}$ from $10^{-2}\rch - 10^2\rch$, with $\Omega$ increasing with $R_{\rm in}$. Note that $\Psi$ is a function of density as well as
ionizing source luminosity. Radiation pressure can be significant in a high density region with relatively lower ionizing luminosity. Note that our full \HII region model grids are incorporated into the dynamical models in Paper II.
}
\end{figure}

For each Cloudy model, we compute the quantities $\Psi$ and $\Omega$, and thus we may think of our models as describing a parameter space $(t, n_{\rm H,in}, \Psi,\Omega)$, as illustrated by Figure \ref{fig:param_example}. This parameter space describes \HII regions for which both radiation and wind pressure run from strong to negligible. To study the effects of winds, we reduce this four-dimensional parameter space to a three-dimensional one by selecting two representative values of $\Omega$: we designate models with $\log\Omega = 2$ as \textit{strong wind} (SW) models, and those with $\log\Omega = -1.5$ as \textit{weak wind} (WW) models. Since our models never produce $\log\Omega = -1.5$ or 2 exactly, we construct these models by interpolation. At each age $t$, density $n_{\rm H,in}$, and luminosity $L$, we find the two models whose values of $\Omega$ bracket our target one, and we compute line luminosities at the target value of $\Omega$ by interpolating between the two bracketing models.

Through this procedure, we obtain a set of four reduced model grids, which we refer to as RPWW (radiation pressure turned on, $\log\Omega = -1.5$), RPSW (radiation pressure turned on, $\log\Omega = 2$), GPWW (radiation pressure turned off, $\log\Omega = -1.5$), and GPSW (radiation pressure turned off, $\log\Omega = 2$).   

It is important to bear in mind that our RP models are physically self-consistent, whereas the GP models are deliberately not.   Thus, RPWW models make a transition from classical spherical Str\"omgren spheres to radiation-confined shells, along the sequence described by 
\citet{Draine11}, as $\Psi$ decreases through unity.   RPSW models are always thin shells: both radiation and wind pressure play a role in confining them, but, as we explain below, wind pressure always dominates.   GPSW models are also thin shells, but due to the neglect of radiation pressure they sample a range of ionization parameters inaccessible to real regions.  Finally, GPWW models are always filled Str\"omgren spheres, even when the radiation force should confine them.  They can also sample unphysically high values of $\calU$.   Our GP models have strictly uniform gas pressure, as they include no other forces.

Each of these model grids gives the line luminosities of \HII regions as a function of the three remaining parameters, $(t,n_{\rm H,in},\Psi)$, or equivalently $(t,n_{\rm H},L)$. We summarize the properties of the models in Table \ref{tab:models}. We will make use of the four full model grids in Paper II, but for the remainder of this paper we concentrate on the particular case $t = 0$, $n_{\rm H,in} = 10$ cm$^{-3}$, in order to understand how the choice of input physics alters the structure of \HII regions. We choose these parameters in particular because they match the ones used by a number of previous authors, thus facilitating easy comparison.

\begin{deluxetable*}{lccc}
\tablewidth{0pt}
\tablecaption{Model Properties}
\tablehead{ \colhead{Model} & \colhead{Acronym} & \colhead{Radiation Pressure?} & \colhead{$\log\Omega$} }
\startdata
Radiation pressure, weak wind & RPWW & yes & $-1.5$ \\
Radiation pressure, strong wind & RPSW & yes & \phs$2$ \\
Gas pressure, weak wind & GPWW & no & $-1.5$ \\
Gas pressure, strong wind & GPSW & no & \phs$2$
\enddata
\label{tab:models}
\end{deluxetable*} 

\section{Results}\label{S:model_results}

\subsection{One-Zone Models}\label{SS:Un_grids}

For a given spectral shape, each ionized parcel with uniform
density and temperature can be characterized by only two parameters:
the density $n_{\rm H}$ and the ionization parameter $\calU =
n_{\gamma,i} / n_{\rm H}$, where $n_{\gamma,i}$ is the number density
of ionizing photons. Therefore there is a unique mapping between
$\calU$ and initial densities on the BPT diagram.
The one-zone models also represent simple analogs of \HII regions, for
one can decompose an \HII region into zones in which $\calU$, $n_{\rm H}$, and ionizing spectrum are nearly constant. 
Thus the one-zone models represent thin, uniformly ionized regions which are very much like the ionized layer of a wind bubble. 
As such, they resemble best the SW models to be discussed in \S3.2.2. 

We run an additional set of Cloudy ``one-zone" calculations in which we only compute the properties of line emission from the first, innermost zone.
In this zone 
we can specify the value of $\calU$ by choosing the the density $n_{\rm H}$ and the bolometric luminosity $L$ (and thus the ionizing photon luminosity $S$). We run models with $n_{\rm H} = 10^{-1} - 10^6$ cm$^{-3}$ in 1 dex steps, and $\calU = 10^{-4} - 10^{0.6}$ in 0.2 dex steps, all using an input spectrum corresponding to our $t=0$ Starburst99 model, and using the same abundances and other parameters as the rest of our models.

\begin{figure}
\plotone{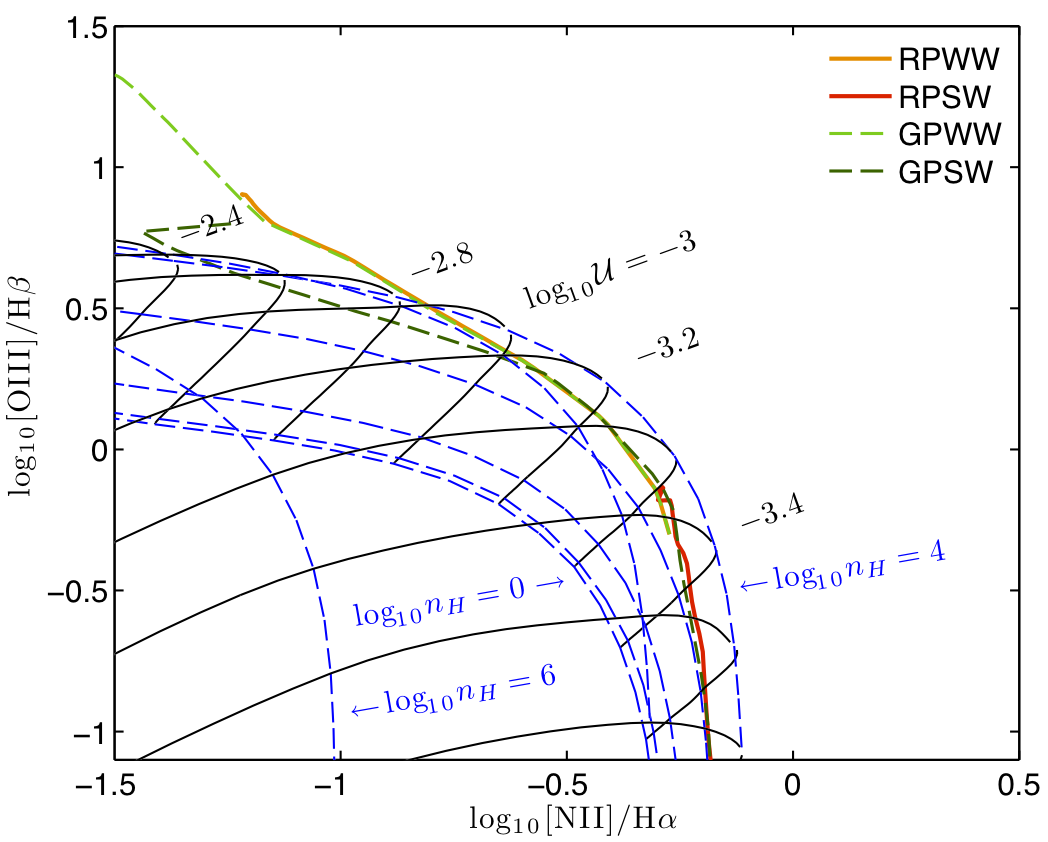}
\caption{
Models in the BPT diagram. Black lines show one-zone models with constant $\calU$, while blue dashed lines show one-zone models of constant $n_{\rm H}$; both are calculated for an ionizing spectrum corresponding to a zero-age stellar population and Solar metallicity. We also show models RPWW, RPSW, GPWW, and GPSW (orange and green lines, as indicated by the legend) with $n_{\rm H,in} = 10$ cm$^{-3}$, calculated with the same ionizing spectrum and metallicity.
\label{fig:BPTonezone}
}
\end{figure}

In Figure~\ref{fig:BPTonezone}, 
we show the constant $\calU$ and constant $n_{\rm H}$ contours marked with black solids lines and blue dashed lines, respectively, on the BPT diagram computed with the one-zone models. 
The ionization parameter $\calU$ and ionizing luminosity $S$ increase from lower right to upper left, and
increasing the density shifts models up and to the right, until the density exceeds $\sim 10^4-10^5$ cm$^{-3}$. 
Beyond this point, the models shift down and to the left, because the density exceeded the critical densities of the [\NII] and [\OIII] emission lines, which
are $6.6 \times 10^4$ cm$^{-3}$ and $6.8 \times 10^5$ cm$^{-3}$, respectively \citep{osterbrock_book}.
In Paper II, we will return to the discussion of line ratios, $\calU$, and critical densities on the BPT diagram, and further discuss 
most extreme \HII regions exceeding the upper limit of line ratios set by \citet{kewley01}, which is based on the mapping between line ratios and $\calU$ but ignored
the effect of densities. 

We note that the BPT locations of macroscopic \HII regions will differ from those of individual gas parcels, because of spatial variations in the physical quantities.  
In all cases $\calU$ drops and the ionizing spectrum changes as one approaches the ionization front, because of selective absorption by neutral H atoms, and in some cases by dust grains.  
When radiation pressure is strong (and is included) and winds are weak, the gas density increases significantly across the layer.  
We therefore anticipate that full \HII region models should differ from the one-zone calculation, even though the innermost zones are accounted by it. Moreover the macroscopic physical parameters, the assumed geometry, and the inclusion or neglect of radiation pressure should affect the BPT loci. We explore these dependencies in the subsequent sections.

\subsection{Full Models}\label{SS:our_models}

We now turn to our four full (radially resolved) \HII region models, RPWW, GPWW, RPSW, and GPSW. In Figure~\ref{fig:BPTonezone}, we overlay these models with $t=0$ and $n_{\rm H,in} = 10$ cm$^{-3}$ on the one-zone calculations. Other choices of density give qualitatively similar results, as long as the density is well below the critical densities of the [\OIII] and [\NII] lines. As with the one-zone models, the full models form a sequence of values defined by 
$\Psi$ or $\calU$, which we control by varying $L$: high-$\Psi$, low-$\calU$, low-$L$ models are found at the bottom right and low-$\Psi$, high-$\calU$, high-$L$ ones at the top left of each sequence.

\subsubsection{Weak Wind Models}
\label{SSS:weak_wind}

When stellar wind pressure is negligible ($\log\Omega = -1.5$), \HII regions at the low $L$ end of the sequence are very similar to each other. This is because $L$ determines the balance between radiation pressure and gas pressure; a high luminosity produces a large $\rch$ (Equation \ref{eq:rch}) and thus a small value of $\Psi$ (Equation \ref{eq:Psi}). Thus when $L$ is low radiation pressure forces are negligible, and the results do not change much depending on whether we include them or not. The density within both \HII regions is roughly constant at $n_{\rm H} = n_{\rm H,in} = 10$ cm$^{-3}$.

\begin{figure}
\plotone{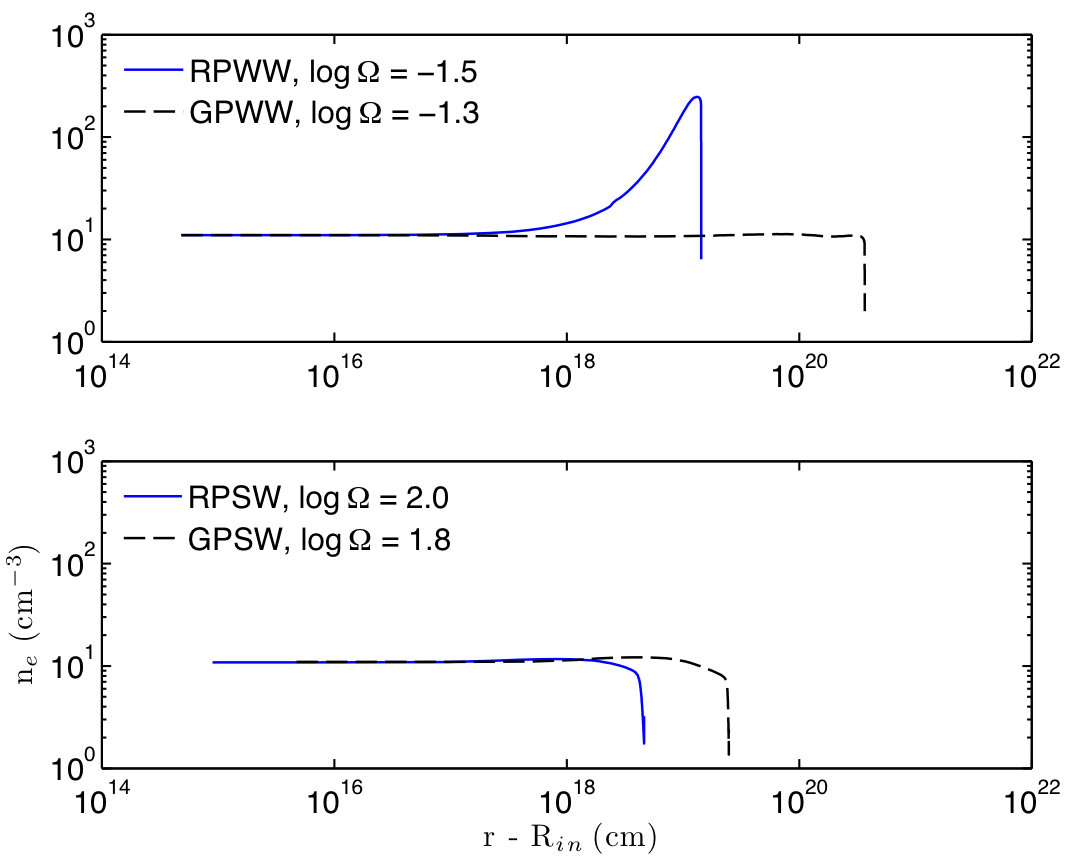}
\caption{Electron density versus radius for sample \HII regions. Top panel: RPWW (blue solid line) and GPWW (black dashed line) regions. Bottom panel: RPSW (blue solid line) and GPSW (black dashed line) regions. The age of the ionizing star cluster in these regions is 0 Myr and the density at the inner boundary is 10 cm$^{-3}$. The luminosity in all models is $10^{43}$ erg s$^{-1}$. We select the value of $R_{\rm in}$ from our grid that gives $\log\Omega$ closest to $-1.5$ and $2$; exact values of $\Omega$ for the four cases shown are as indicated in the legend. See Section~\ref{SS:model_outputs} for details.
Again we note that high luminosity is required here to reach radiation pressure-dominated state because the density is low.
\label{fig:HII_region_models}
}
\end{figure}

At the high $L$, on the other hand, RPWW and GPWW differ substantially. In model GPWW, as $L$ increases, we find that [\OIII]/H$\beta$ increases and [\NII]/H$\alpha$ decreases without limit. In contrast, in model RPWW these line ratios saturate at a finite value. If one were to infer ionization parameters from these line ratios based on one-zone models, one would say that $\calU$ saturates at a finite value in model RPWW, while in model GPWW it can increase without limit as $L$ does. We can understand the difference in behavior by examining the density structures of RPWW and GPWW regions, of which we show an example in Figure \ref{fig:HII_region_models}. At high $L$, RPWW model \HII regions are strongly dominated by radiation pressure. Under force balance, radiation pressure confines ionized gas into a much thinner layer and leads to a steep increase in density towards the IF. Much of the line emission comes from this dense layer, within which $\calU$ is much lower than it is closer to the central source. YM12 discuss this effect in detail. 

This effect does not operate in the GPWW models, where we have artificially disabled radiation pressure. As a result, these \HII regions remain at nearly constant density regardless of the source luminosity. This allows $\calU$ to increase without limit, and in turn allows the  [\OIII]/H$\beta$ and [\NII]/H$\alpha$ line ratios to continue changing even at large $L$.

Finally, it is interesting to note that, despite the uniform pressure in the GPWW models, the actual values of the line ratios are still significantly offset from the corresponding one-zone models of the same density, $n_{\rm H} = 10$ cm$^{-3}$. At small $L$ the shape of the sequence is similar but the models are displaced to slightly higher [\NII]/H$\alpha$ and [\OIII]/H$\beta$, while at large $L$ the deviation is larger and the shape of the sequence is different as well. This difference occurs because, even though the 
pressure 
is uniform in the GPWW models, other quantities are not. In particular, the 
spectrum of the ionizing radiation field varies 
 with radius, due to selective absorption of lower-energy photons 
by neutral H atoms and of higher-energy photons by dust within the ionized layer. 

\subsubsection{Strong Wind Models}
\label{SSS:strong_wind}
 
In the RPSW and GPSW models ($\log\Omega = 2$), strong stellar wind pressure produces large ``voids" of diffuse, high temperature stellar wind gas at the centers of the model \HII regions. As a result, the ionized gas is confined to a thin shell between the wind bubble and the IF.

The location of the RPSW model in the BPT diagram is strikingly far from the locations of other models.
Like the RPWW models, the RPSW models saturate at finite values of [\NII]/H$\alpha$ and [\OIII]/H$\beta$, regardless of how high the luminosity becomes. However, unlike in case RPWW, the saturation values are extraordinarily far down the sequence of one-zone models: $\mbox{[\NII]/H$\alpha$} > 10^{-0.5}$ and $\mbox{[\OIII]/H$\beta$} < 10^0$, corresponding to a one-zone value of $\calU < 10^{-3.3}$. We can understand this effect by considering the relative importance of radiation and wind pressure in controlling the internal structures of \HII regions. A value of $\log\Omega = 2$ requires that $P_{\rm IF} V_{\rm IF}/P_{\rm in} V_{\rm in} = 1.01$. Physically, this amounts to saying that the energy of the wind bubble constitutes 99\% of the internal energy of the entire \HII region. We note that $V_{\rm IF}$ is strictly greater than $V_{\rm in}$. Similarly, $P_{\rm IF}$ is strictly greater than $P_{\rm in}$, since the radiation force necessarily falls to zero at the IF, and thus pressure balance requires that gas pressure at the IF exceed that at the edge of the wind bubble. Thus models with $\log\Omega = 2$ necessarily have both $V_{\rm IF} \approx V_{\rm in}$ and $P_{\rm IF} \approx P_{\rm in}$. This corresponds to the \HII region being a thin shell of nearly constant gas pressure. Figure \ref{fig:HII_region_models} shows an example of this uniform density.

The RPSW configuration clearly cannot have radiation pressure as a significant force.
If the radiation pressure force were significant, then we could not have $P_{\rm IF} \approx P_{\rm in}$, since the pressure at the IF would be pure gas pressure, and this would have to balance the gas plus radiation pressure at the inner edge. The conclusion of this analysis is that it is not possible to construct a physically consistent model 
in which wind pressure and radiation pressure are both strong in the dimensionless sense. 
 Indeed, our model grids reflect this fact in that there are no models with radiation pressure turned on that
are simultaneously dominated by wind ($\Omega\gg1$) and dominated by radiation pressure ($\Psi\ll1$). 
This physical effect manifests in the BPT diagram as a saturation in the range of line ratios that the RPSW models are able to reach. As discussed above, the location of an \HII region driven by a stellar source in the BPT diagram is effectively controlled by $\calU$, the photon to baryon ratio. However, $\calU$ may also be thought of as a measure of the importance of radiation pressure, since increasing the photon number density relative to the baryon density also increases the radiation pressure relative to gas pressure. The fact that RPSW models cannot reach small values of
 $\Psi$ also means that they cannot reach large values of $\calU$, and thus cannot reach the line ratios associated with large $\calU$. 
YM12 used this point to derive upper limits on the wind energy budget within individual \HII regions and entire galaxies.

In contrast, radiation pressure is completely neglected in GPSW models. 
As there are no other forces to balance gas pressure gradients, these models have $P_{\rm IF}=P_{\rm in}$ 
independent of the luminosity,
and thus a value of $\log\Omega = 2$ simply implies that 
the shell is very thin: $ R_{\rm IF}=1.003 R_{\rm in}$. 
In these models one can achieve
arbitrarily high $\calU$
by raising the luminosity and increasing $R_{\rm in}$ to keep up with $R_{\rm IF}$. 
Figure \ref{fig:HII_region_models} shows an example of such a model. Thus the GPSW models are not restricted in the 
range of $\calU$ they can represent. 
However, the comparison with the RPSW models shows that GPSW models at high $\calU$ are unphysical, because 
radiation force would have compressed the gas and limited $\calU$ in a real region. 

The comparison between RPSW, GPSW, and one-zone models shown in Figure \ref{fig:BPTonezone} also reflects these effects. Both RPSW and GPSW models
are wind-dominated and therefore 
 have nearly uniform densities, and thus can be reasonably-well represented by one-zone models, leaving aside the issues of radiation field hardening and non-uniform temperature discussed in Section~\ref{SSS:weak_wind}. 
 Thus both RPSW and GPSW models follow the one-zone sequence reasonably closely.
They differ only in the range of $\calU$ values within that sequence that they are able to sample.

\section{Comparison to Previous Work}\label{sec:model_comparison}

\begin{figure}
\plotone{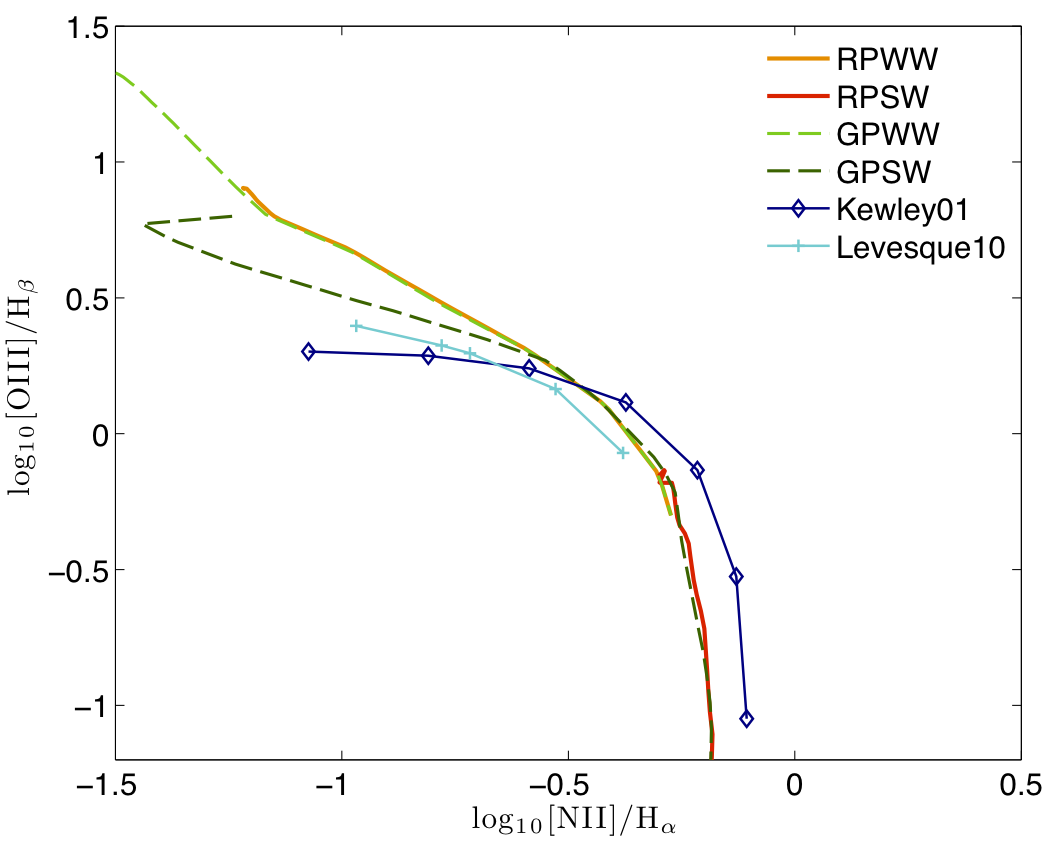}
\caption{
A comparison of model results on the BPT diagram. All results shown are for $n_{\rm H,in} = 10$ cm$^{-3}$ and a spectrum corresponding to a zero-age stellar population. The D00 model (marked as Kewley01 in the legend) 
is shown in dark blue, the L10 model result is in light blue, and our model results are shown in orange and green lines. 
\label{fig:BPT}
}
\end{figure}

It is interesting to revisit earlier published models for \HII region line ratios in the context of our exploration of how these line ratios respond to changes in the included physics. We have chosen our model parameters in Starburst99 and Cloudy to be as close as possible to those used by D00 in order to facilitate this comparison. In Figure~\ref{fig:BPT}, we compare our four model results with the results computed by D00 and L10, at the same age of 
ionizing star cluster (0 Myr) and same density $n_{\rm H,in} = 10$ cm$^{-3}$.

The D00 model, which did not include radiation pressure and adopted plane-parallel ionized gas slabs at a fixed density, is essentially a wind-dominated model. This is because a plane-parallel slab can be thought of as a thin shell of material at roughly fixed distance from the ionizing source,
 and the only way to create a thin shell of constant gas pressure is to confine it with hot gas. 
Therefore our closest model to the D00 model is GPSW, and indeed we find that our GPSW results agree with the D00 model fairly well. Differences in line ratios between D00 and GPSW are around 0.1 to 0.2 dex. Our model sequence extends somewhat further, but this is simply a result of our having used a slightly larger range of input luminosities. The agreement between the models confirms that our 
Cloudy calculations, with input parameters set very close to the D00 settings, 
can reasonably well reproduce the earlier results.
However, we note the comparison shows that the D00 models are not physically realistic at high luminosity, because one cannot neglect radiation pressure in very bright \HII regions. Radiation pressure limits the physical range of $\calU$, particularly for wind-confined slabs, and models without radiation pressure such as those of D00 do not properly capture this effect.

The models from L10 are similar to those of D00 in that they are based on plane-parallel ionized gas slabs, but the L10 models include radiation pressure. Therefore when $\calU$ is low, the regions must be confined by wind pressure (like our RPSW). On the other hand,
when $\calU$ is high, the L10 models should be confined by radiation pressure (like our RPWW). 
However, Figure \ref{fig:BPT} shows that overall L10's models closely track our GPSW curve (maximum separation $<0.1$ dex).
In light of our results, we can see that the L10 models, while not physically inconsistent, do represent a rather odd cut through parameter space. There are {\em two} structural parameters describing \HII regions, and the L10 models sample a one-dimensional path through this two-dimensional space. Along this path the ratio of wind pressure to radiation pressure varies systematically from large values ($\Omega \gg 1$, $\Psi\gg 1$) at low ionization parameter to small values ($\Omega \ll 1$, $\Psi \ll 1$) at high ionization parameter. There is no obvious physical reason such a systematic variation in wind to radiation pressure strength should occur, particularly since the ratio of stellar wind momentum flux to luminosity is roughly the same for all O stars \citep{repolust04a}.

\section{Conclusions}\label{S:BPT_discussion}

We have computed a grid of quasi-static \HII region models using Starburst 99 and Cloudy that covers a large range of density, luminosity, and stellar population age. In order to understand how radiation pressure and stellar winds alter \HII regions' internal structures and observable line emission, we run two sets of models, one with radiation pressure enabled and one with it disabled, and we vary the radius at which the inner, wind-dominated bubble ends and the photoionized region begins. In the manner, we construct four sets of model \HII regions: (1) ones with radiation pressure and weak stellar winds (RPWW), (2) wind bubbles that also include radiation pressure (RPSW), (3) \stromgren spheres where radiation pressure is ignored and winds are weak (GPWW), and (4) wind-dominated bubbles where radiation pressure is disabled (GPSW). We then explore how each set of \HII regions populates the BPT diagram.

Our models reveal a number of interesting effects. All models form a sequence that runs from the lower right corner of the BPT diagram (high [\NII]/H$\alpha$, low [\OIII]/H$\beta$) to the upper left corner (low [\NII]/H$\alpha$, high [\OIII]/H$\beta$), with the position of an \HII region along the sequence dictated by its luminosity, or equivalently its effective ionization parameter $\calU$. However, the range 
of $\calU$ explored by the models
is limited when radiation pressure is included. Because strong radiation pressure, which would produce high $\calU$, 
also causes gas to pile into a dense shell, the characteristic value of $\calU$ within the shell is limited at a finite value. (See YM12 for more detail.)
As a consequence, models which neglect radiation pressure can reflect an arbitrarily high value of $\calU$, which real regions cannot. 

The interaction of winds with radiation pressure further enhances this effect. We show that a stellar wind-dominated region cannot also have strong radiation pressure while remaining in hydrostatic balance, and as a result  
the 
range of $\calU$ is severely limited. This means that wind-dominated \HII regions can never occupy the upper-left portion of the BPT diagram, and, conversely, those \HII regions that are observed to lie in this region must either have negligible wind pressure, be far from pressure balance, or be kept in pressure balance by forces other than gas and radiation pressure (e.g.~strong magnetic pressure; YM12).  The most realistic option, and the one favored by direct observations of nearby \HII regions \citep{harperclark09, lopez11a} as well as mid-infrarared line ratios (YM12) is the first one: wind pressure is not dynamically significant, at least for bright \HII regions.  
Further, the fact that the high-redshift galaxy population has characteristically high ionization parameters implies that radiation pressure is significant within these galaxies' ionized zones, in an ionization-weighted sense.


We 
have compared
 our results to the earlier models of \citet[D00]{dopita00} and \citet[L10]{levesque10}. In these models the \HII region is assumed to be a wind-dominated thin ionized shell, which corresponds to our GPSW model. We find that this model agrees well with the results of D00 and L10. 
However, 
we show that these models are inconsistent at the high luminosity end. The
D00 models neglect radiation pressure for \HII regions where it is non-negligible.
The L10 models include radiation pressure, but we show that the assumed plane-parallel slab geometry is physically realistic only if the strength of the ratio of stellar wind pressure to radiation pressure varies systematically with \HII region properties in a physically unexpected manner.

While these calculations provide insight into how the physics driving \HII regions' structures translates into observable properties such as line ratios, a full model for where galaxies fall in the BPT diagram requires attention to \HII regions' dynamical expansion as well as their internal structure. This problem is the subject of Paper II.

\acknowledgements 
This project was initiated during the 2010 International Summer Institute for Modeling in Astrophysics (ISIMA) summer program, whose support is gratefully acknowledged. 
SCCY and CDM would like to acknowledge an NSERC Discovery grant and conversations with Stephen Ro and Shelley Wright.
MRK acknowledges support from an Alfred P. Sloan Fellowship, from the National Science Foundation through grant CAREER-0955300, 
 from NASA through Astrophysics Theory and Fundamental Physics grant NNX09AK31G, and a Chandra Telescope Grant.

\end{document}